\documentclass{Interspeech2024}

\usepackage{graphicx}
\usepackage{subcaption}
\usepackage{multirow}
\usepackage{adjustbox}
\usepackage{siunitx}

\interspeechcameraready

\title{Tradition or Innovation: A Comparison of Modern ASR Methods for Forced Alignment}

\name[affiliation={1}]{Rotem}{Rousso}
\name[affiliation={1}]{Eyal}{Cohen}
\name[affiliation={1}]{Joseph}{Keshet}
\name[affiliation={2}]{Eleanor}{Chodroff}

\address{
  $^1$Technion - Israel Institute of Technology, Israel\\
  $^2$University of Zurich, Switzerland}
\email{rotem.rousso@campus.technion.ac.il, jkeshet@technion.ac.il, eleanor.chodroff@uzh.ch}


\begin{document}

\maketitle

\begin{abstract}
Forced alignment (FA) plays a key role in speech research through the automatic time alignment of speech signals with corresponding text transcriptions. Despite the move towards end-to-end architectures for speech technology, FA is still dominantly achieved through a classic GMM-HMM acoustic model. This work directly compares alignment performance from leading automatic speech recognition (ASR) methods, WhisperX and Massively Multilingual Speech Recognition (MMS), against a Kaldi-based GMM-HMM system, the Montreal Forced Aligner (MFA). Performance was assessed on the manually aligned TIMIT and Buckeye datasets, with comparisons conducted only on words correctly recognized by WhisperX and MMS. The MFA outperformed both WhisperX and MMS, revealing a shortcoming of modern ASR systems. These findings highlight the need for advancements in forced alignment and emphasize the importance of integrating traditional expertise with modern innovation to foster progress.
\end{abstract}

\section{Introduction}

Forced alignment (FA) is the process of aligning a transcript with the corresponding audio signal to determine the temporal boundaries of units such as words or phones. Such alignment can facilitate downstream processing of the sound file by providing a quick and accurate location of speech units within a longer audio file. 
Accurate labeling and alignment of audio files hold significant potential for advances in both linguistic research and resource development for language communities. In particular, phonetic studies have increasingly relied on this technique, as it greatly expedites acoustic-phonetic analysis of spoken data. 
By some estimates, forced alignment can be as much as 200 to 400 times faster than manual alignment \cite{YuanEtAl2013Automaticphoneticsegmentationusingboundarymodels, YoungMcGarrah2023ForcedalignmentNordiclanguagesRapidlyconstructinghighqualityprototype}.

Historically, FA algorithms have been based on the acoustic model of a modular Automatic Speech Recognition (ASR) system, where the algorithm is \emph{forced} to identify the best path through the acoustic frames given a user-provided sequence of words or phones \cite{brugnara1993automatic}. This results in an alignment of the words or phones to the acoustics. This is in contrast to ASR systems, which aim to predict words or phones from the acoustic frames. 

While traditionally associated with ASR systems, FA is not inherently part of the core recognition process. Instead, it is a critical task within the broader domain of automatic speech processing. This encompasses various tasks related to analyzing and understanding spoken language, including speech recognition, speech synthesis, prosody analysis, and phonetic transcription. While FA has often been conducted using components of ASR systems, including the acoustic model as mentioned above, it is not strictly necessary for ASR and often serves as a preprocessing step. 

In recent years, the progress in ASR technology has been transformative and revolutionary, with remarkable increases in speech recognition ability, particularly from systems like wav2vec 2.0 \cite{baevski2020wav2vec}, HuBERT \cite{hsu2021hubert}, and Whisper \cite{radford2023robust}. Despite the significant advancements in various aspects of speech recognition systems, the classical HMM-GMM algorithm remains one of the leading methods for forced alignment tasks. One of the leading toolkits for implementing this is the Montreal Forced Aligner (MFA) \cite{mcauliffe2017montreal}, which regularly ranks among one of the top forced alignment toolkits \cite{gonzalez2020comparing, mahr2021performance}. The field has also relied extensively on related algorithms and systems, including but not limited to MAUS and WebMAUS \cite{schiel1999automatic, kisler2017multilingual}, easyAlign \cite{goldman2011easyalign}, the Prosodylab-Aligner \cite{gorman2011prosodylab}, FAVE and the Penn Forced Aligner \cite{RosenfelderFave2011, p2fa}, Gentle \cite{ochshorn2017gentle}, and LaBB-CAT \cite{fromont-hay-2012-labb}. There exist other specially designed algorithms for phoneme alignment, such as \cite{keshet2007large} and \cite{hazan2010direct}, that provide very accurate alignment but need to be trained on supervised phoneme-aligned data and are very sensitive to inaccurate lexicons. Recently, Zhu \emph{et al.} \cite{zhu2022phone} proposed two wav2vec 2.0-based models for both text-dependent and text-independent phone-to-audio alignment.

For forced alignment, HMM-based systems have an intuitive advantage over end-to-end systems in two respects: first, HMMs have a direct temporal relationship between acoustic frames and labeled states, and second, words are commonly modeled as a sequence of phones, which directly correspond to a sequence of states. In contrast, modern end-to-end ASR systems are optimized for the direct prediction of characters or tokens and lack fine-grained phonetic representation \cite{graves2006connectionist}, though systems such as WhisperX \cite{bain2023whisperx}, do promote accurate time alignments. 

Our study provides a comparative evaluation of forced alignment as performed by an HMM-based method and modern ASR methods, with a focus on word-level alignment. We evaluate the HMM-based MFA \cite{mcauliffe2017montreal} against two end-to-end systems, the Massively Multilingual Speech Recognition (MMS) system \cite{pratap2023mms} based on wav2vec 2.0 \cite{baevski2020wav2vec} and WhisperX \cite{bain2023whisperx} based on Whisper \cite{radford2023robust}.
The MFA is capable of both phone- and word-level alignment, given the direct modeling of words as a sequence of phones.

For the end-to-end systems, word-level forced alignment was achieved by changing the ASR outputs to match orthographic units with audio segments as they were originally trained to predict characters or tokens.
Alignment was evaluated on two manually aligned English corpora: TIMIT \cite{garofolo1993darpa}, a phone-level transcribed read speech corpus, and Buckeye \cite{raymond2002analysis}, a hand-corrected phone- and word-level corpus of spontaneous English speech.

The structure of the paper is as follows. The next section provides some background on each of the models we evaluate for forced alignment. Section~\ref{sec:method} describes the method used for comparison, including the dataset and the evaluation metrics. The results are presented in Section~\ref{sec:results}, and Section~\ref{sec:discussion} concludes the paper.

\section{Background}\label{sec:backgraund}

In this section, we briefly describe the systems under evaluation in terms of their primary use case, their architecture and training data, and how they can generate forced alignments.

\subsection{Montreal Forced Aligner (MFA)}

The Montreal Forced Aligner (MFA) provides a user-friendly wrapper to the Kaldi ASR toolkit with the primary purpose of developing and deploying acoustic models for phonetic FA\footnote{https://montreal-forced-aligner.readthedocs.io}. The acoustic models have a GMM-HMM architecture and represent the probability of an acoustic sequence given a word sequence; these form a core component of the traditional modular ASR system with separate acoustic and language models. For ASR, the GMM-HMM acoustic model was the dominant approach until the advent of more advanced models with neural architectures or end-to-end modeling \cite{amodei2016deep, collobert2016wav2letter}. The MFA, however, has the primary goal of performing forced alignment and not automatic speech recognition. 

The MFA uses 39 MFCC acoustic features extracted every 10 msec with a processing window of 25 msec. An HMM-based model requires an aligned transcription, which is generated using the Expectation-Maximization (EM) algorithm on the audio file and corresponding transcript.  
The training of the acoustic model has four stages: training monophone models (GMM-HMMs of context-independent phones), training triphone models (GMM-HMMs of context-dependent phones), then applying speaker-adapted refinements including linear discriminant analysis with a maximum likelihood linear transform (LDA+MLLT), and speaker adaptive training (SAT) with feature space maximum linear likelihood regression (fMLLR) \cite{povey2006feature}.

During inference, FA is implemented in the MFA by specifying the audio file, a corresponding transcript, and a pronunciation lexicon which contains a mapping of orthographic words to phonetic transcriptions.
At inference, the orthographic transcription is mapped to a phonetic transcription using a pronunciation lexicon. Then the phonetic transcription is represented as a sequence of HMM triphones. The FA process is implemented via the Viterbi algorithm that identifies the most probable path of acoustic frames given the state sequence. Transitions between HMM states corresponding to different phones or words are then used respectively as phone-level or word-level boundaries in forced alignment. Given the frame shift of 10 msec between frames, the resolution of alignment also corresponds to 10 msec, with a minimal phone duration of 30 msec (as a phone model has a minimum of three states).

\subsection{The Massively Multilingual Speech (MMS) Model}

The MMS FA is based on a single multilingual automatic speech recognition model for 1,107 languages. This model is based on wav2vec 2.0 \cite{baevski2020wav2vec}, a transformer-encoder-based framework for self-supervised learning of speech representations. 

Wav2vec 2.0 operates by first converting the speech signal into latent representations through a multi-layer convolutional neural network (CNN) every 20 msec. These representations are then quantized and used to predict the original sequence in the context of a masked language model, similar to techniques used in NLP for models like BERT \cite{kenton2019bert}. The model is pre-trained on a large corpus of unlabeled audio data, allowing it to learn rich, contextual representations of speech sounds. After pre-training, it can be fine-tuned with a smaller amount of labeled data for specific tasks like speech recognition. For the task of ASR it is trained with the Connectionist Temporal Classification (CTC) loss function \cite{graves2006connectionist}. One major advantage of CTC is that it does not need an exact alignment between the input and output. This means that it is easier to train the system on a large amount of data and to use character-based targets. Nevertheless, the lack of a clear temporal alignment also poses a great challenge when using the system for forced alignment.

\subsection{WhisperX}

Whisper \cite{radford2023robust} is an ASR system trained on 680K hours that is implemented as an encoder-decoder transformer \cite{NIPS2017_3f5ee243}. The speech is converted to an 80-channel log-magnitude Mel spectrogram on windows of 25 msec and a frame shift of 10 msec. The Whisper encoder generates a representation of (up to) 30 seconds of input speech. The decoder gets this representation as input as well as the previously predicted token sequence and outputs the next token. Note that a single representation is used throughout the generation of the predicted text. The encoder-decoder is trained jointly to predict the next token from a set of 50k tokens (words or sub-words) by minimizing the cross-entropy loss function. Whisper performs at or above human listener accuracy \cite{Kim2024Automatic}. However, its architecture and its loss function raise a problem in generating meaningful token alignments \cite{radford2023robust, bain2023whisperx}. 

WhisperX \cite{bain2023whisperx} proposed a technique to improve the predicted word timestamps. Bain \emph{et al.} \cite{bain2023whisperx} proposed to segment the speech into 30 sec chunks using an external voice activity detector (VAD). Following this segmentation, they employ forced phoneme alignment with an external phoneme model to generate word-level timestamps. The phoneme model was based on the character-based wav2vec 2.0 pre-trained base model; however, the details on how the wav2vec 2.0 character-based model is adapted for phoneme alignment are not described in their publication.

\section{Method}\label{sec:method}

In this section, we outline the methodology employed to evaluate the performance of the forced alignment task generated by MFA (both phone and word level), MMS, and Whisper X (word level). We start by describing the datasets and the evaluation metrics used for evaluation.

\subsection{Data}
We selected TIMIT \cite{garofolo1993darpa} and Buckeye \cite{povey2006feature} datasets for their high-quality speech recordings and corresponding phonetic and orthographic timed transcriptions. TIMIT is a corpus of American English read speech with both orthographic and phonetic transcriptions. It contains a total of 630 speakers and 6300 utterances that span 5.4 hours. Our evaluation includes 39,834 words and 177,080 phonemes. The average TIMIT utterance was 3.1 seconds.
The Buckeye Corpus of conversational speech spans 40 hours of hand-transcribed speech from 40 speakers of American English. Our evaluation includes 285,347 words and 858,386 phonemes. The average Buckeye utterance was 531 seconds.

\subsection{Procedure}


We used the MFA acoustic model trained on 982 hours of LibriSpeech \cite{librispeech}\cite{mfa_english_us_arpa_acoustic_2022}. The model's architecture and training procedure is outlined in Section~\ref{sec:backgraund}, and the alignment procedure was configured according to the specific requirements of MFA \cite{chodroff2018corpus}.

The pronunciation lexicon for the MFA English acoustic model used the ARPABET phonetic alphabet. In our evaluation of TIMIT, we mapped the original set of 61 phones provided by TIMIT to a reduced set of 39 phones according to \cite{lee1989speaker}, and further combined the closure with the burst of all stop consonants. The resulting phone set directly corresponded to the required ARPABET phone set. The Buckeye transcriptions already used ARPABET, so no further modification was necessary.  

When comparing the performance of MMS and WhisperX to MFA, it is essential to recognize that the evaluation aims to assess word-level forced alignment using ASR systems that were not specifically designed for the task of FA.

As part of the evaluation process for MMS and WhisperX, it is necessary to accurately match their output words with the corresponding words in the ground truth transcripts. This involved identifying matching words and finding the nearest matches, while disregarding any words that are mislabeled or off by more than 500 msec. 
For TIMIT, which had 39,834 words in the reference transcript, MMS correctly recognized 29,057 words, and WhisperX correctly recognized 37,685 words. For Buckeye, which had 285,347 words in the reference transcript, MMS correctly recognized 259,189 words and WhisperX correctly recognized 278,480 words.

\subsection{Evaluation metrics}

We assessed alignment accuracy for each algorithm across each dataset by examining the difference in the end timestamp between the alignment output and the input, as well as by calculating the percentage of phones or words aligned under a constant threshold. This methodology aligns with previous works in the field \cite{keshet2007large}. We conducted this assessment using various thresholds, chosen based on \cite{mcauliffe2017montreal}. 
The threshold-based evaluation of forced alignment systems has also been conducted in a range of other studies \cite{mahr2021performance, goldman2011easyalign, YuanEtAl2013Automaticphoneticsegmentationusingboundarymodels, YoungMcGarrah2023ForcedalignmentNordiclanguagesRapidlyconstructinghighqualityprototype, DicanioEtAl2013}. 

By adopting the same assessment method as the MFA, we ensure consistency and comparability in our evaluation process. This decision allows for a direct comparison between our results and those obtained in the original MFA study, providing valuable insights into the performance of newer algorithms relative to established ones.
For each algorithm and dataset combination, we further analysed alignment accuracy by examining the mean and median of the total differences for each comparison. In addition, we assessed the F$_1$ score within 20 msec thresholds.

\section{Results}\label{sec:results}

Table~\ref{tab:fa_timit} and Table~\ref{tab:fa_buckeye} present the performance of MFA, MMS, and WhisperX of word alignment on TIMIT and Buckeye, respectively. In these tables, each row represents an alignment model, and each column represents a time resolution threshold. 
To account for the particularly long input utterance durations for Buckeye and high risk of drift in the alignment, we also report the results of alignments correctly placed within the threshold of 500 msec in Table~\ref{tab:fa_buckeye}. 
These results consistently show that the MFA outperforms the MMS and WhisperX timestamps at all thresholds in terms of tolerance values. 

Table~\ref{tab:mean_median} presents the performance of all methods both on TIMIT and on Buckeye in terms of the mean alignment shift, the 
median shift, and the F$_1$ score within 20 msec thresholds. Again it is noticeable that MFA outperforms any other method in terms of alignment accuracy, while MMS and WhisperX are way much more accurate than any HMM-GMM model.
The outcomes further indicate that MFA and WhisperX algorithms perform better on TIMIT than on Buckeye, whereas MMS has more variable performance between the two corpora.

We additionally present the performance of MFA on phone-level alignments. Table~\ref{tab:mfa_phone} represents the performance for TIMIT and Buckeye (rows) for several time resolutions (columns). 
Each number indicates the percentage of correctly placed boundaries of the correct phones within a given resolution. 
We can also see that the phones are aligned more accurately for TIMIT, which is read speech, than for Buckeye, which is conversational speech. The results of Buckeye are somewhat lower than those results published in \cite{mcauliffe2017montreal}, despite the use of a comparable acoustic model. 
The discrepancy is likely due to differences in how the utterances were segmented and input into the MFA for downstream phone- and word-level alignments. To generate the alignment of Buckeye, the input utterances could be as long as several minutes. These input utterances were likely longer than those used in \cite{mcauliffe2017montreal}, which separated utterances at any 150-msec stretch of nonspeech audio.
The long input could have resulted in drift in the alignment prediction, which, in turn, led to a very high mean alignment offset (see also Table~\ref{tab:mean_median}).

The MFA also performs better at the word-level than the phone-level alignment, something that was not observed in the original MFA paper \cite{mcauliffe2017montreal}. This could be due to some word boundaries naturally aligning with pauses in speech and being positioned near periods of quiet; however, it is unclear what led to the discrepancy between the original and current findings with the Buckeye corpus.

\begin{table}[h!]
\renewcommand{\arraystretch}{1.3}
\centering
\caption{TIMIT word-level performance of MFA \cite{mcauliffe2017montreal}, MMS \cite{pratap2023mms}, and WhisperX \cite{bain2023whisperx}. The columns represent correctly detected boundaries at a given resolution in msec.}\label{tab:fa_timit}
\begin{tabular}{lcccc}
\hline
&\multicolumn{4}{c}{\textbf{alignment accuracy [\%]}}\\
\cline{2-5}
 & $t \leq 10$ & $t \leq 25$ & $t \leq 50$ & $t \leq 100$\\
\hline
MFA & 41.6 & 72.8& 89.4& 97.4\\
MMS & 18.6 & 43.5& 75.7& 94.7\\
WhisperX & 22.4 & 52.7&82.4& 94.2\\
\hline
\end{tabular}
\end{table}

\begin{table}[h!]
\renewcommand{\arraystretch}{1.3}
\centering
\caption{Buckeye word-level performance of MFA \cite{mcauliffe2017montreal}, MMS \cite{pratap2023mms}, and WhisperX \cite{bain2023whisperx}. The columns represent correctly detected boundaries at a given resolution in msec. Thresh$^{500}$ indicates alignments correctly placed within the threshold of 500 msec.}\label{tab:fa_buckeye}
\begin{adjustbox}{max width=0.48\textwidth}
\begin{tabular}{lccccc}
\hline
 &  & \multicolumn{4}{c}{\textbf{alignment accuracy [\%]}} \\
\cline{3-6}
 &  \textbf{Thresh$^{500}$}  & $t \leq 10$ & $t \leq 25$ & $t \leq 50$ & $t \leq 100$  \\
\hline
MFA & - &  39.8 & 69.9& 84.9& 91.8\\
MMS & -& 25.0 & 52.7 &75.0& 87.9\\
WhisperX & -& 18.8 & 43.1&67.4& 77.4\\
\hline

MFA & +& 41.1 & 72.2& 87.6& 94.8\\
MMS & +& 25.8 & 54.2 &77.2& 90.5\\
WhisperX & +& 22.8 & 52.3&81.8& 93.9\\
\hline
\end{tabular}
\end{adjustbox}
\end{table}

\begin{table}[h!]
\renewcommand{\arraystretch}{1.3}
\centering
\caption{Performance in terms of mean and median (msec), as well as F-score assessed at a threshold of 20 msec ($\text{F}_1^{20}$) for the phone-level and word-level alignments from the MFA\cite{mcauliffe2017montreal}, word-level alignments from MMS \cite{pratap2023mms}, and word-level alignments from WhisperX \cite{bain2023whisperx}.}\label{tab:mean_median}
\begin{adjustbox}{max width=0.48\textwidth}
\begin{tabular}{llccSSS}
\hline
\textbf{Method}  & \textbf{Dataset}& \textbf{Thresh$^{500}$} &  \textbf{Level} &\textbf{Mean}& \textbf{Med}&  \textbf{$\text{F}_1^{20}$}\\
\hline
MFA & TIMIT & - & phon &  133.4& 12.5& 66.0\\
MFA & Buckeye & - & phon &  1085.9& 15.9& 56.2\\

MFA & TIMIT & - & word &  21.9&12.5& 65.7\\
MMS & TIMIT & - & word & 68.5& 29.3& 35.4\\
WhisperX & TIMIT & - & word & 34.3& 23.5& 43.5\\

MFA & Buckeye & - & word &  976.5& 13.6& 63.4\\
MMS & Buckeye & - & word & 208.3& 23.1& 45.0\\
WhisperX & Buckeye & - & word &11685.3& 30.1 & 35.6\\

MFA & Buckeye & + & word &   27.8& 12.9& 65.4\\
MMS & Buckeye & + & word &41.0& 22.2& 46.3\\
WhisperX & Buckeye & + & word & 36.4& 23.7& 43.3\\
\hline
\end{tabular}
\end{adjustbox}
\end{table}

\begin{table}[h!]
\renewcommand{\arraystretch}{1.3}
\centering
\caption{Phone-level performance of MFA \cite{mcauliffe2017montreal}. The columns represent correctly detected boundaries at a given resolution in msec.}\label{tab:mfa_phone}
\begin{tabular}{lcccc}
\hline
& \multicolumn{4}{c}{\textbf{alignment accuracy [\%]}}\\
\cline{2-5}
& $t \leq 10$  & $t \leq 25$ & $t \leq 50$ & $t \leq 100$ \\
\hline
TIMIT & 38.6& 72.3& 81.1& 84.6\\
Buckeye & 35.3& 60.6& 68.9& 72.7\\
\hline
\end{tabular}
\end{table}

\section{Discussion}\label{sec:discussion}

Our evaluation revealed that MFA outperforms MMS and WhisperX in the comparison of word-level alignments at all time resolutions. For speech researchers, the classical GMM-HMM architecture still outperforms simple adaptations of the modern end-to-end ASR systems for forced alignment.
Indeed, the HMM-based system of the MFA has a high temporal resolution of 10 msec relative to MMS and WhisperX that operate over longer stretches of audio. WhisperX does improve the temporal resolution of words and utterances over Whisper's relatively low performance, with the stated goal of improved forced alignment \cite{bain2023whisperx}. Nevertheless, our findings demonstrate that the HMM-based system is still preferable for forced alignment tasks. 

One of the major bottlenecks to analyses that depend on forced alignment is the ability to obtain an accurate transcription of the text. \cite{ahnoutlier}. Even if an orthographic transcription is obtained, this still needs to be converted to a phonetic transcription that is then usable with a pretrained acoustic model, or sufficient data must be present to train an acoustic model that performs well with the specified phone set. In terms of speech recognition, both MMS and WhisperX outperform any HMM-based model in terms of word error rate \cite{baevski2020wav2vec, radford2023robust}. Systems such as MMS and WhisperX will likely play a valuable role in the pipeline towards forced alignment in generating usable transcripts or revising existing transcripts that may contain errors.

This contribution of transcript generation or correction could serve as a valuable component in improving forced alignment. Indeed, it complements another method that has demonstrated to significantly improve forced alignment, namely ``recursive forced alignment''. Recursive forced alignment refers to a multi-stage forced alignment procedure, where the first-pass alignment is used to identify shorter utterance-level boundaries. The utterance-level boundaries can then be used to constrain the forced alignment input domain (i.e., the input utterance duration), which results in overall more accurate word- and phone-level boundaries.\cite{moreno1998recursive, GonzalezEtAl2018Recursiveforcedalignmenttestminoritylanguage, BarthEtAl2020Usingforcedalignmentsociophoneticresearchminoritylanguage}

The end-to-end models likely had poorer alignment due to their architecture and training procedure. The MMS model is a transformer encoder trained using contrastive loss function in a self-supervised manner to predict a masked 20-msec speech frame. 
To turn this model into an ASR system, a linear layer is added, and it is trained to predict characters using the CTC loss function. 
As mentioned earlier, the CTC loss removes the need for pre-aligned training data and considers the network outputs as a probability distribution over all possible alignments. 
The lack of pre-aligned training data, of course, leads to poor alignment at inference.

Whisper, on the other hand, is an encoder-decoder transformer, which represents the input speech utterance as a whole. 
The decoder does not work at the speech-frame level but is trained to predict the next token, given the representation and the previous sequence of the predicted tokens using the cross-entropy (multi-class) loss function. 
This mechanism may lose the alignment of the tokens within the encoder representation and the depth of the decoder.

Finally, the phone-level performance from the MFA demonstrates that the GMM-HMM architecture straightforwardly yields such alignments. Many phoneticians and speech researchers rely on the phone-level transcription to understand phonetic and phonological variation across talkers and languages. A phone-level representation is not straightforwardly available in end-to-end ASR systems such as MMS and WhisperX, the field is actively working to develop this area \cite{xu2021simple, zhu2022phone}. A future direction for the present study is to investigate how to use modern algorithms and architectures with good temporal resolution for phone-level forced alignment. 


The findings here indicate that despite the considerable advances in speech recognition using end-to-end systems, traditional GMM-HMM architectures appear to be optimal for forced alignment tasks, at least in the current state of writing. This paper serves as a call to action for additional research and development of deep learning algorithms specifically designed for forced alignment tasks.

\bibliographystyle{IEEEtran}
\bibliography{mybib}

\begin{thebibliography}{10}
\providecommand{\url}[1]{#1}
\csname url@samestyle\endcsname
\providecommand{\newblock}{\relax}
\providecommand{\bibinfo}[2]{#2}
\providecommand{\BIBentrySTDinterwordspacing}{\spaceskip=0pt\relax}
\providecommand{\BIBentryALTinterwordstretchfactor}{4}
\providecommand{\BIBentryALTinterwordspacing}{\spaceskip=\fontdimen2\font plus
\BIBentryALTinterwordstretchfactor\fontdimen3\font minus \fontdimen4\font\relax}
\providecommand{\BIBforeignlanguage}[2]{{%
\expandafter\ifx\csname l@#1\endcsname\relax
\typeout{** WARNING: IEEEtran.bst: No hyphenation pattern has been}%
\typeout{** loaded for the language `#1'. Using the pattern for}%
\typeout{** the default language instead.}%
\else
\language=\csname l@#1\endcsname
\fi
#2}}
\providecommand{\BIBdecl}{\relax}
\BIBdecl

\bibitem{YuanEtAl2013Automaticphoneticsegmentationusingboundarymodels}
J.~Yuan, N.~Ryant, M.~Liberman, A.~Stolcke, V.~Mitra, and W.~Wang, ``Automatic phonetic segmentation using boundary models,'' in \emph{Proceedings of the {{Annual Conference}} of the {{International Speech Communication Association}}, {{INTERSPEECH}}}, 2013, pp. 2306--2310.

\bibitem{YoungMcGarrah2023ForcedalignmentNordiclanguagesRapidlyconstructinghighqualityprototype}
N.~J. Young and M.~McGarrah, ``Forced alignment for {{Nordic}} languages: {{Rapidly}} constructing a high-quality prototype,'' \emph{Nordic Journal of Linguistics}, vol.~46, no.~1, pp. 105--131, May 2023.

\bibitem{brugnara1993automatic}
F.~Brugnara, D.~Falavigna, and M.~Omologo, ``Automatic segmentation and labeling of speech based on hidden markov models,'' \emph{Speech Communication}, vol.~12, no.~4, pp. 357--370, 1993.

\bibitem{baevski2020wav2vec}
A.~Baevski, Y.~Zhou, A.~Mohamed, and M.~Auli, ``wav2vec 2.0: A framework for self-supervised learning of speech representations,'' \emph{Advances in neural information processing systems}, vol.~33, pp. 12\,449--12\,460, 2020.

\bibitem{hsu2021hubert}
W.-N. Hsu, B.~Bolte, Y.-H.~H. Tsai, K.~Lakhotia, R.~Salakhutdinov, and A.~Mohamed, ``Hubert: Self-supervised speech representation learning by masked prediction of hidden units,'' \emph{IEEE/ACM Transactions on Audio, Speech, and Language Processing}, vol.~29, pp. 3451--3460, 2021.

\bibitem{radford2023robust}
A.~Radford, J.~W. Kim, T.~Xu, G.~Brockman, C.~McLeavey, and I.~Sutskever, ``Robust speech recognition via large-scale weak supervision,'' in \emph{International Conference on Machine Learning}.\hskip 1em plus 0.5em minus 0.4em\relax PMLR, 2023, pp. 28\,492--28\,518.

\bibitem{mcauliffe2017montreal}
M.~McAuliffe, M.~Socolof, S.~Mihuc, M.~Wagner, and M.~Sonderegger, ``Montreal forced aligner: Trainable text-speech alignment using kaldi.'' in \emph{Interspeech}, vol. 2017, 2017, pp. 498--502.

\bibitem{gonzalez2020comparing}
S.~Gonzalez, J.~Grama, and C.~E. Travis, ``Comparing the performance of forced aligners used in sociophonetic research,'' \emph{Linguistics Vanguard}, vol.~6, no.~1, p. 20190058, 2020.

\bibitem{mahr2021performance}
T.~J. Mahr, V.~Berisha, K.~Kawabata, J.~Liss, and K.~C. Hustad, ``Performance of forced-alignment algorithms on children's speech,'' \emph{Journal of Speech, Language, and Hearing Research}, vol.~64, no.~6S, pp. 2213--2222, 2021.

\bibitem{schiel1999automatic}
F.~Schiel, ``Automatic phonetic transcription of non-prompted speech,'' in \emph{Proceedings of the 14th {{International Congress}} on {{Phonetic Sciences}} ({{ICPhS}})}, 1999, pp. 607--610.

\bibitem{kisler2017multilingual}
T.~Kisler, U.~Reichel, and F.~Schiel, ``Multilingual processing of speech via web services,'' \emph{Computer Speech \& Language}, vol.~45, pp. 326--347, Sep. 2017.

\bibitem{goldman2011easyalign}
J.-P. Goldman, ``{EasyAlign}: an automatic phonetic alignment tool under praat,'' in \emph{Twelfth Annual Conference of the International Speech Communication Association (Interspeech)}, 2011.

\bibitem{gorman2011prosodylab}
K.~Gorman, J.~Howell, and M.~Wagner, ``{Prosodylab-aligner}: A tool for forced alignment of laboratory speech,'' \emph{Canadian Acoustics}, vol.~39, no.~3, pp. 192--193, 2011.

\bibitem{RosenfelderFave2011}
\BIBentryALTinterwordspacing
I.~Rosenfelder, J.~Fruehwald, K.~Evanini, S.~Seyfarth, C.~Brickhouse, K.~Gorman, H.~Prichard, and J.~Yuan, ``{{FAVE}}: {{Forced}} alignment and vowel extraction,'' Zenodo, Aug. 2022. [Online]. Available: \url{https://zenodo.org/record/593309}
\BIBentrySTDinterwordspacing

\bibitem{p2fa}
J.~Yuan and M.~Liberman, ``Speaker identification on the {{SCOTUS}} corpus,'' in \emph{Proceedings of {{Acoustics}}}, 2008, pp. 5687--5690.

\bibitem{ochshorn2017gentle}
\BIBentryALTinterwordspacing
{\relax RM}.~Ochshorn and M.~Hawkins, ``Gentle forced aligner,'' 2017. [Online]. Available: \url{github.com/lowerquality/gentle}
\BIBentrySTDinterwordspacing

\bibitem{fromont-hay-2012-labb}
\BIBentryALTinterwordspacing
R.~Fromont and J.~Hay, ``{{LaBB-CAT}}: an annotation store,'' in \emph{Proceedings of the australasian language technology association workshop 2012}, P.~Cook and S.~Nowson, Eds., {Dunedin, New Zealand}, Dec. 2012, pp. 113--117. [Online]. Available: \url{https://aclanthology.org/U12-1015}
\BIBentrySTDinterwordspacing

\bibitem{keshet2007large}
J.~Keshet, S.~Shalev-Shwartz, Y.~Singer, and D.~Chazan, ``A large margin algorithm for speech-to-phoneme and music-to-score alignment,'' \emph{IEEE Transactions on Audio, Speech, and Language Processing}, vol.~15, no.~8, pp. 2373--2382, 2007.

\bibitem{hazan2010direct}
T.~Hazan, J.~Keshet, and D.~McAllester, ``Direct loss minimization for structured prediction,'' \emph{Advances in neural information processing systems}, vol.~23, 2010.

\bibitem{zhu2022phone}
J.~Zhu, C.~Zhang, and D.~Jurgens, ``Phone-to-audio alignment without text: A semi-supervised approach,'' in \emph{ICASSP 2022-2022 IEEE International Conference on Acoustics, Speech and Signal Processing (ICASSP)}.\hskip 1em plus 0.5em minus 0.4em\relax IEEE, 2022, pp. 8167--8171.

\bibitem{graves2006connectionist}
A.~Graves, S.~Fern{\'a}ndez, F.~Gomez, and J.~Schmidhuber, ``Connectionist temporal classification: labelling unsegmented sequence data with recurrent neural networks,'' in \emph{Proceedings of the 23rd international conference on Machine learning}, 2006, pp. 369--376.

\bibitem{bain2023whisperx}
M.~Bain, J.~Huh, T.~Han, and A.~Zisserman, ``Whisperx: Time-accurate speech transcription of long-form audio,'' in \emph{Proc. {INTERSPEECH} 2023 -- 24th Annual Conference of the International Speech Communication Association}, 2023.

\bibitem{pratap2023mms}
V.~Pratap, A.~Tjandra, B.~Shi, P.~Tomasello, A.~Babu, S.~Kundu, A.~Elkahky, Z.~Ni, A.~Vyas, M.~Fazel-Zarandi \emph{et~al.}, ``Scaling speech technology to 1,000+ languages,'' \emph{arXiv preprint arXiv:2305.13516}, 2023.

\bibitem{garofolo1993darpa}
J.~S. Garofolo, L.~F. Lamel, W.~M. Fisher, J.~G. Fiscus, and D.~S. Pallett, ``Darpa timit acoustic-phonetic continous speech corpus cd-rom. nist speech disc 1-1.1,'' \emph{NASA STI/Recon technical report n}, vol.~93, p. 27403, 1993.

\bibitem{raymond2002analysis}
W.~D. Raymond, M.~Pitt, K.~Johnson, E.~Hume, M.~Makashay, R.~Dautricourt, and C.~Hilts, ``An analysis of transcription consistency in spontaneous speech from the buckeye corpus,'' in \emph{Seventh International Conference on Spoken Language Processing}, 2002.

\bibitem{amodei2016deep}
D.~Amodei, S.~Ananthanarayanan, R.~Anubhai, J.~Bai, E.~Battenberg, C.~Case, J.~Casper, B.~Catanzaro, Q.~Cheng, G.~Chen \emph{et~al.}, ``Deep speech 2: End-to-end speech recognition in english and mandarin,'' in \emph{International conference on machine learning}.\hskip 1em plus 0.5em minus 0.4em\relax PMLR, 2016, pp. 173--182.

\bibitem{collobert2016wav2letter}
R.~Collobert, C.~Puhrsch, and G.~Synnaeve, ``Wav2letter: an end-to-end convnet-based speech recognition system,'' \emph{arXiv preprint arXiv:1609.03193}, 2016.

\bibitem{povey2006feature}
D.~Povey and G.~Saon, ``Feature and model space speaker adaptation with full covariance {G}aussians.'' in \emph{Interspeech}, 2006, pp. 1145--1148.

\bibitem{kenton2019bert}
J.~D. M.-W.~C. Kenton and L.~K. Toutanova, ``Bert: Pre-training of deep bidirectional transformers for language understanding,'' in \emph{Proceedings of NAACL-HLT}, 2019, pp. 4171--4186.

\bibitem{NIPS2017_3f5ee243}
A.~Vaswani, N.~Shazeer, N.~Parmar, J.~Uszkoreit, L.~Jones, A.~N. Gomez, L.~u. Kaiser, and I.~Polosukhin, ``Attention is all you need,'' in \emph{Advances in Neural Information Processing Systems}, vol.~30, 2017.

\bibitem{Kim2024Automatic}
S.-E. Kim, B.~R. Chernyak, O.~Seleznova, J.~Keshet, M.~Goldrick, and A.~R. Bradlow, ``Automatic recognition of second language speech-in-noise,'' \emph{Journal of the Acoustical Society of America}, 2024.

\bibitem{librispeech}
V.~Panayotov, G.~Chen, D.~Povey, and S.~Khudanpur, ``Librispeech: An {ASR} corpus based on public domain audio books,'' in \emph{2015 IEEE International Conference on Acoustics, Speech and Signal Processing (ICASSP)}, 2015, pp. 5206--5210.

\bibitem{mfa_english_us_arpa_acoustic_2022}
M.~McAuliffe and M.~Sonderegger, ``{English (US) ARPA acoustic model v2.0.0a},'' \url{https://mfa-models.readthedocs.io/acoustic/English/English (US) ARPA acoustic model v2_0_0a.html}, Tech. Rep., May 2022.

\bibitem{chodroff2018corpus}
E.~Chodroff, ``Corpus phonetics tutorial,'' \emph{arXiv preprint arXiv:1811.05553}, 2018.

\bibitem{lee1989speaker}
K.-F. Lee and H.-W. Hon, ``Speaker-independent phone recognition using hidden markov models,'' \emph{IEEE Transactions on Acoustics, Speech, and Signal Processing}, vol.~37, no.~11, pp. 1641--1648, 1989.

\bibitem{DicanioEtAl2013}
C.~DiCanio, H.~Nam, D.~H. Whalen, H.~Timothy~Bunnell, J.~D. Amith, and R.~C. García, ``Using automatic alignment to analyze endangered language data: {{Testing}} the viability of untrained alignment,'' \emph{The Journal of the Acoustical Society of America}, vol. 134, no.~3, pp. 2235--2246, Sep. 2013.

\bibitem{ahnoutlier}
E.~P. Ahn, G.-A. Levow, R.~A. Wright, and E.~Chodroff, ``An outlier analysis of vowel formants from a corpus phonetics pipeline,'' in \emph{Proc. {INTERSPEECH} 2023 -- 24\textsuperscript{th} Annual Conference of the International Speech Communication Association}, {Dublin, Ireland}, {Sep.} 2023.

\bibitem{moreno1998recursive}
P.~J. Moreno, C.~F. Joerg, J.-M. Van~Thong, and O.~Glickman, ``A recursive algorithm for the forced alignment of very long audio segments,'' in \emph{International {{Conference}} on {{Spoken Language Processing}} ({{ICSLP}})}, vol.~98, 1998, pp. 2711--2714.

\bibitem{GonzalezEtAl2018Recursiveforcedalignmenttestminoritylanguage}
S.~Gonzalez, C.~Travis, J.~Grama, D.~Barth, and S.~Ananthanarayan, ``Recursive forced alignment: {{A}} test on a minority language,'' in \emph{Proceedings of the 17th {{Australasian International Conference}} on {{Speech Science}} and {{Technology}}}, 2018, pp. 145--148.

\bibitem{BarthEtAl2020Usingforcedalignmentsociophoneticresearchminoritylanguage}
D.~Barth, J.~Grama, S.~Gonzalez, and C.~Travis, ``Using forced alignment for sociophonetic research on a minority language,'' in \emph{University of {{Pennsylvania Working Papers}} in {{Linguistics}}}, vol.~25, 2020, p.~2.

\bibitem{xu2021simple}
Q.~Xu, A.~Baevski, and M.~Auli, ``Simple and effective zero-shot cross-lingual phoneme recognition,'' \emph{arXiv preprint arXiv:2109.11680}, 2021.

\end{thebibliography}

\end{document}